\documentclass[preprint,preprintnumbers,amsmath,amssymb]{revtex4-1}
\usepackage{graphicx}% Include figure files
\usepackage{dcolumn}% Align table columns on decimal point
\usepackage{bm}% bold math
\usepackage{amssymb,amsfonts,amsmath}
\usepackage{graphicx}
\usepackage{threeparttable}
\usepackage{float}
\newcommand{\specialcell}[2][c]{%
  \begin{tabular}[#1]{@{}c@{}}#2\end{tabular}}
  
\begin{document}
%\linenumbers
%\setpagewiselinenumbers
%\modulolinenumbers[5]
%\linenumbers

%\preprint{APS/123-QED}

\title{\bf{Improper magnetic ferroelectricity of nearly pure electronic nature in cycloidal spiral CaMn$_{7}$O$_{12}$} \\[11pt] }

\author{Jin Soo Lim, Diomedes Saldana-Greco, and Andrew M. Rappe}

\affiliation{
  The Makineni Theoretical Laboratories, Department of Chemistry,
  University of Pennsylvania, Philadelphia, PA 19104-6323, USA\\ }

\date{\today}% It is always \today, today,


\begin{abstract}
The noncollinear cycloidal magnetic order breaks the inversion symmetry in CaMn$_{7}$O$_{12}$, generating one of the largest spin-orbit driven ferroelectric polarizations measured to date.  In this Letter, the microscopic origin of the polarization, including its direction, charge density redistribution, magnetic exchange interactions, and its coupling to the spin helicity, is explored via first principles calculations.  The Berry phase computed polarization exhibits almost pure electronic behavior, as the Mn displacements are negligible, $\approx$~0.7~m\textrm{\AA}.  The polarization magnitude and direction are both determined by the Mn spin current, where the \emph{p}-\emph{d} orbital mixing is driven by the inequivalent exchange interactions within the \emph{B}-site Mn cycloidal spiral chains along each Cartesian direction.  We employ the generalized spin-current model with Heisenberg-exchange Dzyaloshinskii-Moriya interaction energetics to provide insight into the underlying physics of this spin-driven polarization. Persistent electronic polarization induced by helical spin order in nearly inversion-symmetric ionic crystal lattices suggests opportunities for ultrafast magnetoelectric response.
\end{abstract}

%add something about the magnetic interactions, less strong message on abstract

\maketitle

\begin{figure}
\centering
\includegraphics[width=0.7\textwidth, angle = 90]{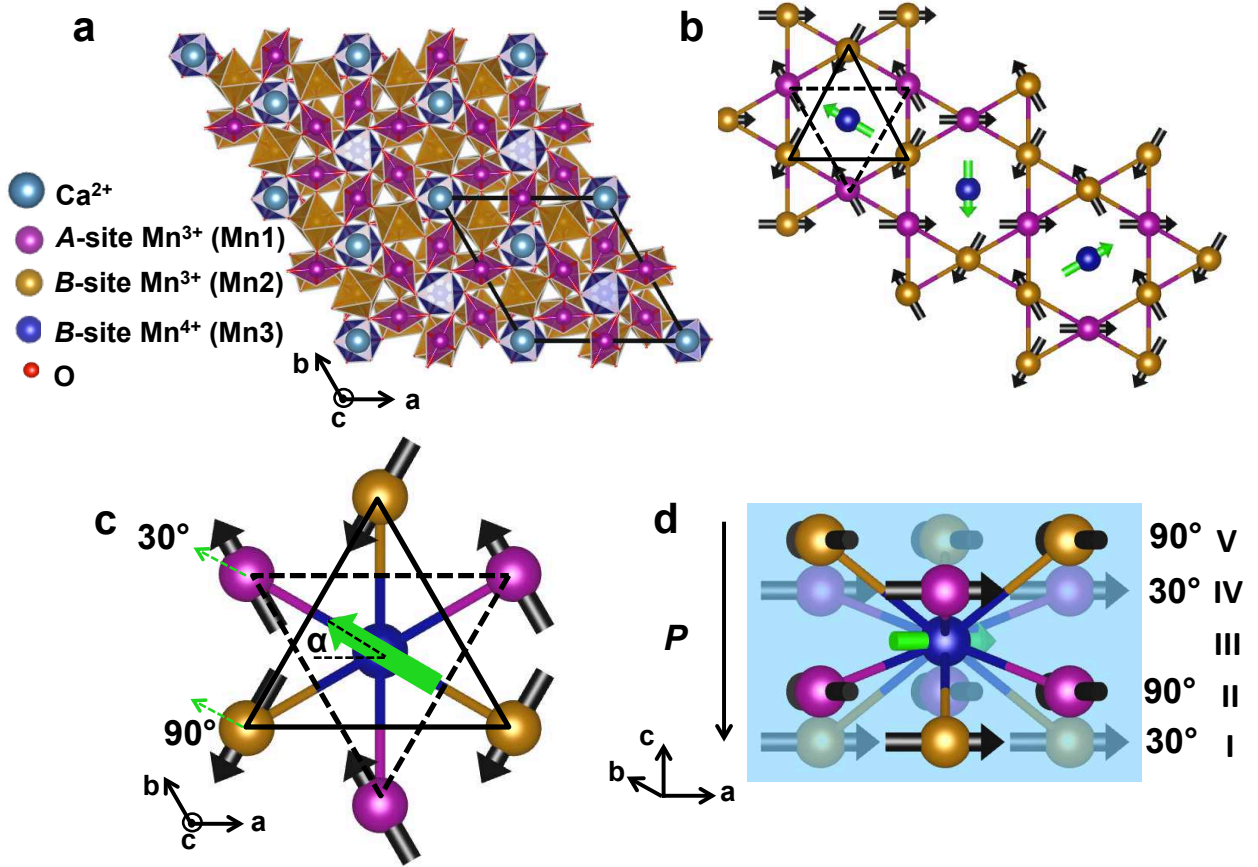}
\caption{(a) The crystal structure of the rhombohedral (\textit{R}$\overline{3}$) phase.  Mn1 (purple, square planar) and Mn2 (gold, octahedral) alternate along parallel \textit{c}-chains, of which sets of three form a hexagonal Kagome lattice.  The hexagonal center is occupied by Ca$^{2+}$ (light blue) and Mn3 (blue, octahedral) alternating along the \textit{c}-axis.  The unit cell is denoted with bold black lines.  (b) Noncollinear magnetic structure with Mn1 and Mn2 spins represented as black arrows and Mn3 spins as green arrows.  Black lines indicate the local hexagonal environment of Mn3 surrounded by Mn1 and Mn2, with solid lines closer to the viewer than dashed lines.  (c) Zoom-in view of the local hexagonal environment in (b).  Mn3 spin configuration is represented by $\alpha$, where $\alpha$ = 30$^\circ$ for (30$^\circ$, 90$^\circ$) configuration with respect to the neighboring (Mn1, Mn2) spins. (d) Side view of (c) showing the local layered structure of five Mn planes (I-V), magnetically inducing net electronic polarization along the \textit{c}-axis.  The blue plane is parallel to the \textit{c}-axis and cuts through the central Mn3, such that the atoms farther away from the viewer are shaded by the plane.}
\label{Fig.1}
\end{figure}

Multiferroics, simultaneously displaying ferroelectricity and intrinsic magnetic ordering, have gained much attention due to the complex physics underlying the magnetoelectric effect and its potential applications in spin-driven electronics~\cite{Spaldin05p391, Fiebig05pR123}.  Based on the nature of the order parameter coupling, multiferroics are classified into type-I and type-II~\cite{Khomskii09p20}.  Type-I consists of 6\textit{s}$^{2}$ lone-pair proper ferroelectrics, as in BiFeO$_{3}$~\cite{Chu08p478, Lebeugle09p257601, Catalan09p2463} and improper ferroelectrics of electronic~\cite{van_den_Brink08p434217} and geometric origins~\cite{van_Aken04p164} including hybrid improper ferroelectrics~\cite{Benedek11p107204}, where ferroelectricity remains largely independent of magnetism.  Type-II essentially refers to improper \textit{magnetic} ferroelectrics where spiral magnetic ordering breaks inversion symmetry, resulting in ionic and/or electronic displacements that provide macroscopic polarization.  Numerous examples include: (a) cycloidal spiral systems including orthorhombic \textit{R}MnO$_{3}$ (\textit{R} = Tb, Dy, Tm)~\cite{Kenzelmann05p087206, Kimura03p55, Goto04p257201, Xiang08p037209, Malashevich08p037210, Pomjakushin09p043019}, CoCr$_{2}$O$_{4}$~\cite{Yamasaki06p207204}, and MnWO$_{4}$~\cite{Taniguchi06p097203} , (b) triangular-lattice systems with proper screw-type spiral~\cite{Arima07p073702}, such as RbFe(MoO$_{4}$)$_{2}$~\cite{Kenzelmann07p267205}, CuFeO$_{2}$~\cite{Kimura06p220401}, and \textit{A}CrO$_{2}$ (\textit{A} = Cu, Ag, Li, Na)~\cite{Kimura06p220401, Seki08p067204, Kimura08p140401}, and (c) exchange-striction systems with collinear magnetism, such as Ca$_{3}$(CoMn)O$_{6}$~\cite{Choi08p047601}, orthorhombic \textit{R}MnO$_{3}$ (\textit{R} = Ho-Lu, Y)~\cite{Sergienko06p227204}, DyFeO$_{3}$~\cite{Tokunaga08p097205}, and Ni$_{3}$V$_{2}$O$_{8}$~\cite{Lawes05p087205}. Despite their relatively small ferroelectric polarization and low Curie temperature, type-II multiferroics are of tremendous technological relevance, potentially leading to the design of robust room-temperature multiferroics with large spontaneous polarization and ultrafast switchability.  In order to achieve this, theoretical insight into spin-induced polarization mechanisms is necessary.
Three microscopic mechanisms have been proposed to explain the emergence of ferroelectricity \textbf{\textit{P}} in spin-spiral multiferroics~\cite{Cheong07p13, Wang09p321, Tokura14p076501}.  First, the exchange striction model proposes that the symmetric exchange interaction in a $\uparrow$ $\uparrow$ $\downarrow$ $\downarrow$ spin order causes ferromagnetically coupled ions to move toward each other, generating \textbf{\textit{P}}$_{12}$~$\propto$~\textbf{\textit{e}}$_{12}$~(\textbf{\textit{S}}$_{1}$~$\cdot$~\textbf{\textit{S}}$_{2}$)~\cite{Tokura14p076501}.  Here, \textbf{\textit{P}}$_{12}$ is the local polarization induced by the interaction between the two neighboring spin sites 1 and 2, \textbf{\textit{S}}$_{1}$ and \textbf{\textit{S}}$_{2}$ are the vector spins on the respective sites, and \textbf{\textit{e}}$_{12}$ is a unit vector connecting the two magnetic ions.  Second, two analytically equivalent scenarios exist within the spin-current (KNB) model~\cite{Katsura05p057205}, where \textbf{\textit{P}}$_{12}$~$\propto$~\textbf{\textit{e}}$_{12}$~$\times$~(\textbf{\textit{S}}$_{1}$~$\times$~\textbf{\textit{S}}$_{2}$) describes: (a) a nonmagnetic anion moving in response to the Dzyaloshinskii-Moriya (DM) interaction between the two canted spin sites (inverse-DM interaction)~\cite{Sergienko06p227204}; (b) electronic charge distribution shifting in response to the spin-current, defined as \textbf{\textit{j}}$_{\textrm{s}}$~=~\textbf{\textit{S}}$_{1}$~$\times$~\textbf{\textit{S}}$_{2}$~\cite{Katsura05p057205}.  Third, the spin-dependent \textit{p}-\textit{d} hybridization model arising from spin-orbit coupling (SOC) causes an intrasite polarization along the metal-ligand bond indicated by \textbf{\textit{P}}$_{\rm ml}$~$\propto$~(\textbf{\textit{S}}$_{\rm m}$~$\cdot$~\textbf{\textit{e}}$_{\rm ml}$)$^2$~\textbf{\textit{e}}$_{\rm ml}$~\cite{Kimura06p220401, Murakawa12p174106, Jia06p224444}, where \textbf{\textit{e}}$_{\rm ml}$ is the metal-ligand unit vector.
Recently, CaMn$_{7}$O$_{12}$ manifested one of the largest magnetically-induced ferroelectric polarizations measured to date (\textit{P}~=~2870~$\mu$C/m$^2$)~\cite{Johnson12p067201}.  Microscopic mechanisms involving the three models discussed above~\cite{Tokura14p076501} have been proposed: exchange striction and DM interaction~\cite{Lu12p187204, Zhang13p075127, Cao15p064422}, inverse-DM interaction~\cite{Perks12p1}, and spin-dependent \textit{p}-\textit{d} hybridization~\cite{Zhang13p075127}.  However, a unified picture that explains the direction of the polarization, the charge density redistribution, and the role of ionic displacements is still needed.
Here, we report on the ferroelectric polarization of nearly pure electronic nature in CaMn$_{7}$O$_{12}$ induced by its noncollinear cycloidal magnetic ground state, computed via density functional theory (DFT) calculations.  For simplicity and clarity, we preserve inversion symmetry on the ionic lattice while the charge density distribution is permitted to respond to the symmetry-breaking spin pattern; these changes to orbital mixing make the dominant contribution to the polarization.  Theoretically, we employ the generalized spin-current model~\cite{Xiang11p157202} with Heisenberg-exchange DM-interaction energetics to explain both the direction of the electronic polarization and the dependence of its magnitude on spin helicity. 
This quadruple perovskite belongs to the [\textit{A}$\textit{A}'_{3}$][\textit{B}$_{4}$][O$_{12}$] family: [CaMn$_{3}$][Mn$_{4}$][O$_{12}$].  CaMn$_{7}$O$_{12}$ undergoes structural and metal-insulator transition accompanied by charge-ordering at \textit{T} = 440~K with a large change in resistivity at ultrafast time scales~\cite{Troyanchuk98p14903, Huon15p142901}.  The \textit{B}-site Mn ions order into Mn$^{3+}$ and Mn$^{4+}$ with a 3:1 ratio in a centrosymmetric rhombohedral (\textit{R}$\overline{3}$) crystal structure [Fig.\ \ref{Fig.1}(a)], such that the formula is rewritten as [CaMn$_{3}$$^{3+}$][Mn$_{3}$$^{3+}$Mn$^{4+}$][O$_{12}$].  Throughout this Letter, \textit{A}-site Mn$^{3+}$ is designated as Mn1, \textit{B}-site Mn$^{3+}$ as Mn2, and \textit{B}-site Mn$^{4+}$ as Mn3.
The material exhibits two magnetic phase transitions at N$\acute{\textrm{e}}$el temperatures, \textit{T}$_{\textrm{N1}}$~=~90~K and \textit{T}$_{\textrm{N2}}$~=~48~K.  Neutron diffraction measurements demonstrated a noncollinear spin configuration of long-range ordering with propagation vector (0, 0, 1.037) between 48~K and 90~K.  Below 48~K, magnetic modulation with two propagation vectors (0, 0, 0.958) and (0, 0, 1.120) was proposed~\cite{Johnson12p067201}.
All spins lie in the \textit{ab}-plane, and Mn ions along the same \textit{c}-chain or of the same Mn-type and \textit{c}-axis height have identical spin directions.  Magnetic interactions among Mn1 and Mn2 ~\cite{Lu12p187204, Perks12p1} cause spin frustration, causing all spin pairs in adjacent \emph{c}-chains to be 120$^\circ$ from each other [Fig.\ \ref{Fig.1}(b)].  The central Mn3 spin direction is determined by the neighboring Mn ions, three Mn1 and three Mn2.  It has been proposed that the Mn3 adopts a spin direction that is (30$^\circ$, 90$^\circ$)~\cite{Johnson12p067201, Lu12p187204, Dai15p135} or (60$^\circ$, 60$^\circ$)~\cite{Przenioslo13p27, Zhang13p075127} with respect to the surrounding (Mn1, Mn2) spin directions.  The Mn3 spin configuration is conveniently represented by the quantity $\alpha$, where $\alpha$~=~0$^\circ$ for (60$^\circ$, 60$^\circ$) and 30$^\circ$ for (30$^\circ$, 90$^\circ$) [Fig.\ \ref{Fig.1}(c)].  The sign of $\alpha$ indicates the spin helicity and chirality.  The local structure of the hexagonal channel consists of five equidistant \textit{ab}-planes (I-V) repeating periodically along the \textit{c}-axis, where the central layer consists of a single Mn3 [Fig.~\ref{Fig.1}(d)].  The ferroelectric phase transition temperature of the material coincides with the N$\acute{\textrm{e}}$el temperature, \textit{T}$_{\textrm{C}}$ = \textit{T}$_{\textrm{N1}}$ = 90~K, suggesting that the ferroelectricity is spin-driven~\cite{Zhang11p174413, Johnson12p067201}.  The macroscopic polarization is along the \textit{c}-axis ([111] in the pseudocubic coordinates), parallel to the spin helicity vector and perpendicular to the spin rotation plane (\textit{ab}-plane).

We evaluate the commensurate, unmodulated, noncollinear magnetic ground state using the PBEsol~\cite{Perdew08p136406} functional with Hubbard \textit{U} and \textit{J} (Coulomb repulsion and exchange parameter) treated separately and explicitly defined within the rotationally invariant scheme~\cite{Lieuchtenstein95pR5467, Himmetoglu11p115108} along with SOC as implemented in the {\sc Quantum Espresso}~\cite{Giannozzi09p395502} package.  It has been demonstrated that the Hubbard \textit{J} parameter plays a central role in correctly describing noncollinear magnetic systems~\cite{Bousquet10p220402}.   All atoms are represented by norm-conserving, optimized~\cite{Rappe09p1227}, designed nonlocal~\cite{Ramer99p12471} pseudopotentials generated with the {\sc opium} package~\cite{Opium}, including the spin-orbit interaction~\cite{Theurich01p073106} as well as nonlinear core-valence interaction in the Mn pseudopotential via the partial-core correction scheme~\cite{Louie92p1738, Fuchs99p67, Porezag99p14132}.  The Brillouin zone is sampled using a $3\times3\times5$ Monkhorst-Pack~\cite{Monkhorst76p5188} $k$-point mesh.
The energetics and spin direction of collinear and noncollinear magnetic configurations are used to justify the values \textit{U}~=~2~eV and \textit{J}~=~1.4~eV used in our DFT calculations [See Section I, Supplemental Material].  These values are close to those used in previous studies~\cite{Lu12p187204, Zhang13p075127, Cao15p064422, Dai15p135}.  We use the experimental centrosymmetric unmodulated ionic lattice structure~\cite{Slawinski13p392}.  Starting from multiple perturbations of the experimental noncollinear magnetic structure~\cite{Johnson12p067201}, our DFT+\textit{U}+\textit{J}+SOC spin and electronic relaxation shows that the Mn1 and Mn2 spin directions are $\approx$~120$^\circ$ apart, and the Mn3 spin direction converges to $\alpha$~$\approx$~30$^\circ$, \emph{i.e}.\ (30$^\circ$, 90$^\circ$) configuration [Fig.\ \ref{Fig.1}(b)].  If the Mn3 spins are started at $\alpha$~=~0$^\circ$, \emph{i.e}.\ (60$^\circ$, 60$^\circ$), they remain in that symmetry, showing that $\alpha$~=~0$^\circ$ is higher in energy by 3~meV per formula unit.
The relationship between $\alpha$ and the electric polarization \textit{P} is explored by computing \textit{P} through the Berry phase method~\cite{King-Smith93p1651} with and without SOC at different $\alpha$ values [See Section II, Supplemental Material]. The polarization is along the \textit{c}-axis. The most relevant scenarios where $\alpha$~$\approx$~0$^\circ$ or $\approx$~30$^\circ$ are shown in Table I.  Simultaneous ionic relaxation [See Section III, Supplemental Material] gives Mn3 displacement of 0.7~m\textrm{\AA} with total \textit{P}~=~-2900~$\mu$C/m$^2$, in good agreement with the experimental value of 2870~$\mu$C/m$^2$~\cite{Johnson12p067201}.  The ionic displacement is negligible relative to thermal motion at \textit{T}$_{\textrm{C}}$~=~90~K, and it contributes 30$\%$ of the total polarization.  At $\alpha$~$\approx$~0$^\circ$, the polarization vanishes, in agreement with the previous theoretical studies of this system~\cite{Lu12p187204, Zhang13p075127}.  Upon inverting the spin helicity by changing the sign of $\alpha$, the direction of the polarization reverses with the same magnitude.  This is in agreement with the phenomenological ferrroaxial coupling proposed by Johnson \textit{et al.}~\cite{Johnson12p067201}

\begingroup
\begin{table}
\caption{Berry phase computed \textit{P} and converged Mn3 spin direction $\alpha$ with DFT+\textit{U}+\textit{J} with and without SOC.}
\begin{ruledtabular}
\centering
\begin{tabular}{lccc}%
		&	$\alpha$		&	\textit{P} ($\mu$C/m$^2$)	\\ \hline
NSOC	&	0.05$^\circ$	&	0							\\
NSOC	&	30.02$^\circ$	&	-1935						\\
SOC	&	29.30$^\circ$	&	-2030						\\ 
\end{tabular}
\end{ruledtabular}
\end{table}
\endgroup

The nonzero Berry phase polarization for $\alpha$~$\approx$~30$^\circ$ shows that the inversion symmetry is broken, even though the ionic lattice structure is fixed to be centrosymmetric.  The material exhibits cycloidal spiral magnetism along \textit{B}-site Mn2-Mn3 chains in each of the Cartesian [100], [010], and [001] directions.  Upon magnetic inversion symmetry operation, the chain system with $\alpha$~=~0$^\circ$ is unchanged, whereas the one with $\alpha$~=~30$^\circ$ has two out of the three inversion-related Mn3 spin pairs altered [See Section IV, Supplemental Material].  Therefore, only the $\alpha$~=~30$^\circ$ configuration breaks inversion symmetry and generates nonzero ferroelectric polarization, consistent with our calculations. 

\begin{figure}
\centering
\includegraphics[width=0.3\textwidth, angle = 90]{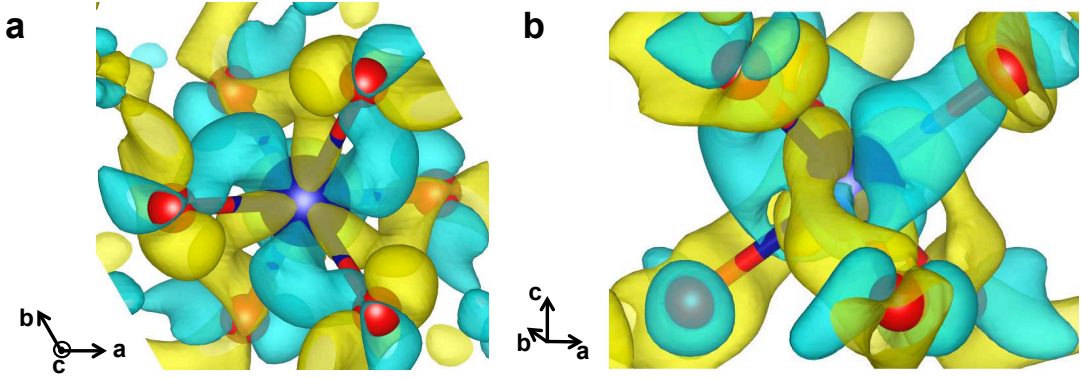}
\caption{Charge density redistribution along the Mn3-O bonds as the magnetic structure changes from $\alpha$ = 0$^\circ$ to 30$^\circ$.  The charge density shift is purely electronic, induced by the change in Mn3 magnetic moment.  (a) Top-view and (b) side-view of the charge density differential.  The top three Mn3-O bonds gain electron density (cyan), while the bottom three Mn3-O bonds lose electron density (yellow).}
\label{Fig.2}
\end{figure}

We compute the charge density redistribution as the magnetic structure goes from $\alpha$~=~0$^\circ$ to 30$^\circ$, $\Delta$$\rho$(\textbf{r})~=~$\rho$(\textbf{r})$_{\alpha=30^{\circ}}$ $-$ $\rho$(\textbf{r})$_{\alpha = 0^{\circ}}$ [Fig.\ \ref{Fig.2}].  The reduced and enhanced charge density isosurfaces reveal that the ferroelectric polarization is localized along the Mn3-O bonds on each local Cartesian direction.  As discussed above, the ions do respond to the charge density redistribution but only by 0.7~m\textrm{\AA}, lowering the energy by only 0.04~meV per formula unit and providing a small contribution to the polarization [See Section III, Supplemental Material].  This suggests that the magnetically-induced ferroelectricity in the system is nearly pure electronic in nature.  However, previously proposed mechanisms have strongly relied on ionic displacements~\cite{Perks12p1, Zhang13p075127} without isolating the electronic contribution.

The spin-current model has been regarded as inapplicable to CaMn$_{7}$O$_{12}$, as the mechanism requires the polarization to lie on the spin-rotation plane [\textbf{\textit{P}}$_{12}$~$\propto$~\textbf{\textit{e}}$_{12}$~$\times$~(\textbf{\textit{S}}$_{1}$~$\times$~\textbf{\textit{S}}$_{2}$), which lies on the \textit{ab}-plane in CaMn$_{7}$O$_{12}$, not along the \textit{c}-axis as observed].  However, Xiang \textit{et al.}~\cite{Xiang11p157202, Xiang13p054404} proposed a generalized spin-current model to analytically explain ferroelectricity induced by spiral magnetism.  Polarization induced by a noncentrosymmetric spin dimer \textbf{\textit{S}$_{1}$} and \textbf{\textit{S}$_{2}$} is written as
\begin{eqnarray}
{P}_{12}^{k} = \frac{1}{2}\sum_{ijl}\Gamma^{ijk}[\boldsymbol{S}_{1} \times \boldsymbol{S}_{2}]^{l} \epsilon_{ijl}
\end{eqnarray}
\noindent where Cartesian coordinates are denoted by \textit{i}, \textit{j}, \textit{k}, and \textit{l}, $\epsilon_{ijl}$ is the Levi-Civita symbol, and $\Gamma^{ijk}$ is a rank-three magnetoelectric coupling tensor with its elements indicating the intersite vector polarizations associated with \textbf{\textit{S}$_{1}$} and \textbf{\textit{S}$_{2}$}. For example, \boldsymbol{$\Gamma$$^{ij}$} is the vector polarization arising from the \textit{i}-component of \textbf{\textit{S}$_{1}$} and the \textit{j}-component of \textbf{\textit{S}$_{2}$}.  Spin inversion requires that \boldsymbol{$\Gamma$$^{ij}$}~=~$-$ \boldsymbol{$\Gamma$$^{ji}$} and consequently \boldsymbol{$\Gamma$$^{ii}$}~=~\textbf{0}, thereby eliminating the diagonal terms in the tensor $\Gamma^{ijk}$ and reducing it to a 3 $\times$ 3 magnetoelectric coupling matrix written in the form:
\begin{eqnarray}
\boldsymbol{\Lambda}^{k}_{l} = \sum_{ij}\Gamma^{ijk} \epsilon_{ijl},
\end{eqnarray}
leading to the following polarization expression [See Section V, Supplemental Material]:
\begin{eqnarray}
\boldsymbol{P}_{12} = \boldsymbol{\Lambda} (\boldsymbol{S}_{1} \times \boldsymbol{S}_{2}).
\end{eqnarray}
\indent It is important to emphasize the dependence of the polarization on the spin current (\textbf{\textit{S}$_{1}$}~$\times$~\textbf{\textit{S}$_{2}$}), rather than the dot product (\textbf{\textit{S}$_{1}$}~$\cdot$~\textbf{\textit{S}$_{2}$}) as in the exchange striction model ~\cite{Lu12p187204, Cao15p064422}.  Exchange striction model results when rotational invariance is assumed by neglecting spin-orbit coupling, which eliminates the nondiagonal terms in the tensor \textbf{$\Gamma^{ijk}$} and generates the dot product [See Section V, Supplemental Material].  However, ferroelectricity in CaMn$_{7}$O$_{12}$ is not rotationally invariant, as global rotation of spins affects the polarization.  We therefore conclude that the generalized spin-current model is more appropriate for CaMn$_{7}$O$_{12}$ in the context of noncollinear spins.

We consider the local hexagonal structure from [Fig.\ \ref{Fig.1}(d)] and the six Mn2-O-Mn3 spin dimer interactions within the cyloidal spiral chains along which the charge redistribution is localized.  The Mn3 spin is designated as \textbf{\textit{S}}$_{\textrm{Mn3}}$, whereas the Mn2 spins of layer I are designated as \textbf{\textit{S}}$_{\textrm{I}}$ and Mn2 spins of layer V as \textbf{\textit{S}}$_{\textrm{V}}$.  The expression for the polarization in terms of $\alpha$ becomes:
\begin{eqnarray}
\boldsymbol{P} &=& 3\boldsymbol{\Lambda}(\boldsymbol{S}_{\textrm{Mn3}} \times \boldsymbol{S}_{\textrm{V}})+3\boldsymbol{\Lambda}(\boldsymbol{S}_{\textrm{Mn3}} \times \boldsymbol{S}_{\textrm{I}}) \nonumber \\
&=& 3\boldsymbol{\Lambda} \boldsymbol{e}_{z} \sin(60^{\circ}+\alpha)-3\boldsymbol{\Lambda} \boldsymbol{e}_{z} \sin(60^{\circ}-\alpha) \nonumber \\
&=& 3\boldsymbol{\Gamma^{ij}} \sin(\alpha).
\end{eqnarray}
This polarization as a function of $\sin$($\alpha$) accounts for its dependence on Mn3 spin direction, \textbf{\textit{P}}($\alpha$~=~0$^\circ$)~=~\textbf{0}, and its coupling to the spin helicity, \textbf{\textit{P}}($-$$\alpha$)~=~$-$\textbf{\textit{P}}($\alpha$). From the above analysis, it is evident that the spin-current (\textbf{\textit{S}$_{1}$}~$\times$~\textbf{\textit{S}$_{2}$}) takes into account both the magnitude and the direction of the polarization in CaMn$_{7}$O$_{12}$.  However, understanding the underlying physics requires further analysis of the intersite magnetic interactions:
\begin{eqnarray}
E_{\textrm{12}} &=& E_{\textrm{SE}}+E_{\textrm{DM}} \nonumber \\
&=& J_{\textrm{12}}(\boldsymbol{S}_{\textrm{1}} \cdot \boldsymbol{S}_{\textrm{2}})+\boldsymbol{D}_{\textrm{12}} \cdot (\boldsymbol{S}_{\textrm{1}} \times \boldsymbol{S}_{\textrm{2}}).
\end{eqnarray}
The first term is the Heisenberg symmetric exchange energy (\textit{E}$_{\textrm{SE}}$), and the second term is the DM antisymmetric exchange energy (\textit{E}$_{\textrm{DM}}$).  \textit{J}$_{\textrm{12}}$ is the exchange coupling between magnetic sites 1 and 2, and the DM vector is defined as \textbf{\textit{D}}$_{12}$~$\propto$~\textbf{\textit{r}}$_{1}$~$\times$~\textbf{\textit{r}}$_{2}$, where \textbf{\textit{r}}$_{1}$ and \textbf{\textit{r}}$_{2}$ are vectors connecting each metal to the intersite ligand.  Considering the same six spin dimer interactions, the \emph{total} magnetic interaction energy becomes [See Section VI, Supplemental Material]:
\begin{eqnarray}
E &=& E_{\textrm{Mn3-V}} + E_{\textrm{Mn3-I}} \nonumber \\
&=& 3[J\cos(\alpha)-D^{z}\sin(\alpha)].
\end{eqnarray}
\indent Because we use the commensurate, unmodulated structure without orbital-ordering, \textit{J}$_{\textrm{Mn3-V}}$~=~\textit{J}$_{\textrm{Mn3-I}}$~=~\textit{J}~\cite{Perks12p1, Cao15p064422}.  Furthermore, \textit{J}~$<$~0~\cite{Lu12p187204} because (a) the alternation of filled and empty \textit{x}$^{\textrm{2}}$-\textit{y}$^{\textrm{2}}$ orbitals on Mn2 and Mn3 along the cycloidal spiral chain promotes ferromagnetic exchange, and (b) the large deviation of Mn3-O-Mn2 bond angles from 180$^{\circ}$ weakens antiferromagnetic interactions~\cite{Perks12p1}.  Additionally, the DM vectors for both Mn3-V and Mn3-I interactions have a $-$\textbf{\textit{c}} component with a magnitude of \textit{D}$^{z}$.

The minimum of the total energy in Eq.\ (6) directly depends on the strength of the magnetic interactions.  Setting $\frac{d\emph{E}}{d\alpha}$=0 leads to $\alpha_{\textrm{min}}$=$\tan^{-1}$(-$\frac{\textit{D}}{\textit{J}}$).  Previous DFT calculations reported $\mid$\textit{D}/\textit{J}$\mid$ $\approx$ 0.54 in CaMn$_{7}$O$_{12}$~\cite{Lu12p187204}, indicating unusually strong Mn3-Mn2 DM interaction compared to other magnetic insulators where $\mid$\textit{D}/\textit{J}$\mid$~$\lesssim$~0.1 is usually expected~\cite{Moriya60p91}.  Considering the reported ratio and the interacting nature, \textit{J}~$<$~0 and \textit{D}$^{z}$~$>$~0, the DM interaction lowers the total magnetic interaction energy by shifting $\alpha$ from 0$^{\circ}$ to 30$^{\circ}$, such that \textit{E}($\alpha$~=~30$^{\circ}$)~$<$~\textit{E}($\alpha$~=~0$^{\circ}$), consistent with our results.

\begingroup
\begin{table}
\caption{Energetics of the magnetic interactions for Mn3-V and Mn3-I with $\alpha$ = 0$^{\circ}$ and 30$^{\circ}$. }
\begin{ruledtabular}
\centering
\begin{tabular}{lccccc}
                                     &                                                  & \emph{E}                    & \emph{E}$_{\textrm{SE}}$     & \emph{E}$_{\textrm{DM}}$ \\ \hline
$\alpha$ = 0$^{\circ}$   &  \specialcell{Mn3-V \\ Mn3-I}    & 3\emph{J}                  &  $\frac{3}{2}$\emph{J}             &    $\frac{3\sqrt{3}}{2}$\emph{D}$^{z}$ \\ \hline
$\alpha$ = 30$^{\circ}$  &  \specialcell{Mn3-V \\ Mn3-I}  & $\frac{3}{2}$($\sqrt{3}$\emph{J}-\emph{D}$^{z}$)    &   \specialcell{ 0 \\ $\frac{3\sqrt{3}}{2}$\emph{J}}    &\specialcell{3\emph{D}$^{z}$ \\ $\frac{3}{2}$\emph{D}$^{z}$} \\
\end{tabular}
\end{ruledtabular}
\end{table}
\endgroup

It is well-known that DM interaction favors noncollinear magnetism in an otherwise collinear magnetic order, thereby inducing a weak local ferromagnetic behavior in an antiferromagnet~\cite{Moriya60p91}.  The energetics analysis in Table II shows that the DM interaction, together with the symmetric exchange interaction, favors $\alpha$~=~30$^{\circ}$ over $\alpha$~=~0$^{\circ}$.  This causes Mn3-V interaction (90$^{\circ}$ alignment) to be inequivalent to Mn3-I interaction (30$^{\circ}$ alignment).  Smaller spin alignment is associated with larger \textit{E}$_{\textrm{DM}}$ and smaller \textit{E}$_{\textrm{SE}}$.  Within the context of symmetric exchange, Mn3-I interaction (30$^{\circ}$ alignment) can be understood as more of ferromagnetic double-exchange character than Mn3-V interaction (90$^{\circ}$ alignment), which is more of antiferromagnetic superexchange character.  The difference leads to a weak exchange striction of nonionic character, where the electrons are slightly more localized in Mn3-V regime than in Mn3-I regime, consistent with our observed charge density redistribution along the Mn3-O bonds [Fig.\ \ref{Fig.2}].  The three Mn3-O bonds pointing toward layer V (with +\textbf{\textit{c}} components) gain electron density, whereas those pointing toward layer I (with -\textbf{\textit{c}} components) lose electron density, thereby generating a net polarization along -\textbf{\textit{c}} direction.

\begin{figure}
\centering
\includegraphics[width=0.7\textwidth, angle = 90]{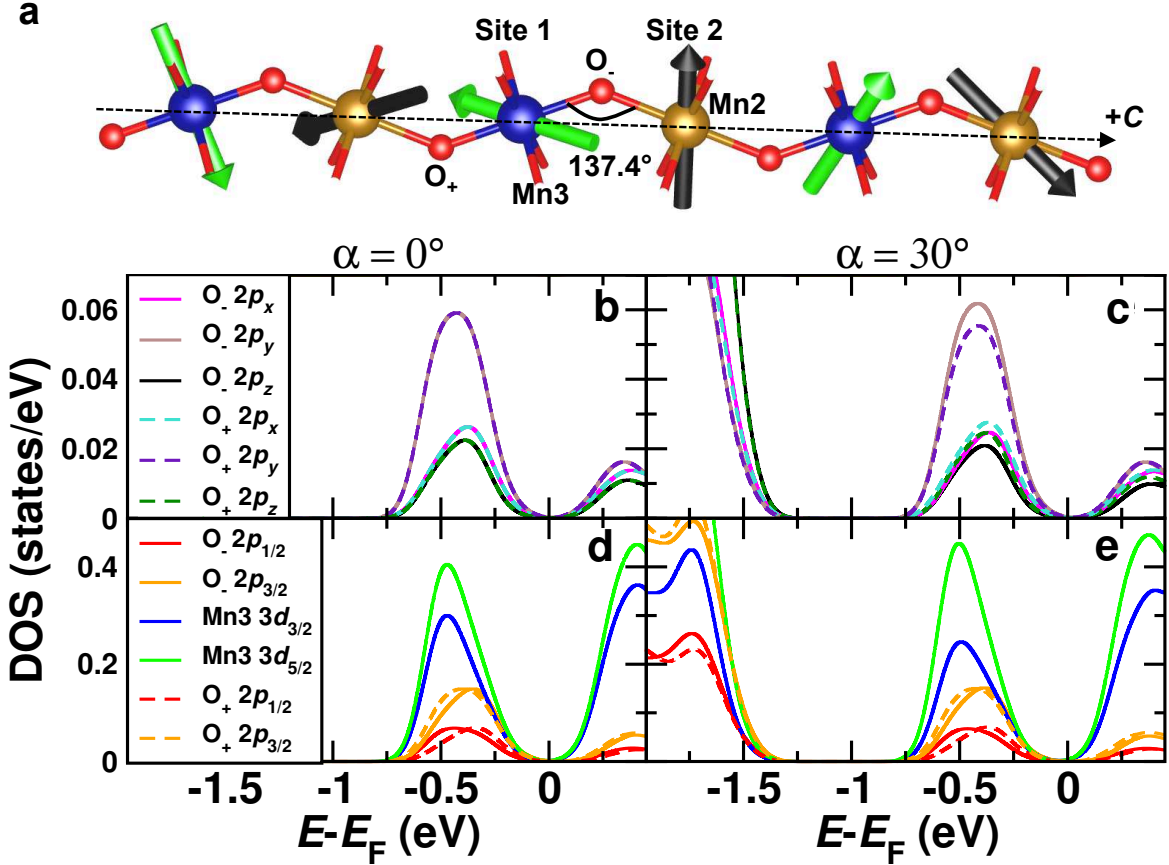}
\caption{(a) Mn2-O-Mn3 chain forms a zig-zag pattern with a bond angle of 137.4$^\circ$.  Orbital-projected density of states for all 2\emph{p}-orbital subshells of O$_{+}$ (solid lines) and O$_{\text{--}}$ (dashed lines) when (b) $\alpha$~=~0$^\circ$ and (c) $\alpha$~=~30$^\circ$.  The spin-orbit coupled states for Mn3 and surrounding O atoms when (d) $\alpha$~=~0$^\circ$ and (e) $\alpha$~=~30$^\circ$.  The charge density redistribution along the Mn3-O bonds is evidenced by the changes in the orbital mixing O$_{+}$ vs. O$_{-}$when $\alpha$~=~30$^\circ$.}
\label{Fig.3}
\end{figure}

The effect of inequivalent exchange interaction on the charge density distribution is manifested in orbital mixing.  We analyze the orbital-projected density of states (PDOS) along the O$_{+}$---Mn3---O$_{\text{--}}$---Mn2 chain [Fig.\ \ref{Fig.3}a].  O$_{+}$ and O$_{\text{--}}$ refer to the oxygens along the reduced and enhanced charge density bonds, respectively.  The \emph{p}-\emph{d} mixing is observed primarily between Mn3~3\emph{d} and O~2\emph{p}.  The total 2\emph{p} PDOS (not shown) exhibits no difference between $\alpha$~=~0$^\circ$ and $\alpha$~$\approx$~30$^\circ$.  However, a significant difference arises within the \emph{p}$_{x}$, \emph{p}$_{y}$, and \emph{p}$_{z}$ orbitals [Fig.\ \ref{Fig.3}b-c].  The 2\emph{p} orbitals of O$_{+}$ and O$_{\text{--}}$ show the same density when $\alpha$~=~0$^\circ$.  However, once the Mn3 spins break the inversion symmetry at $\alpha$~$\approx$~30$^\circ$ and the ferroelectric polarization emerges, the 2\emph{p} orbital densities of O$_{+}$ and O$_{\text{--}}$ become inequivalent.  The orbital mixing between Mn3~3\emph{d} and O$_{-}$~2\emph{p} is slightly enhanced merely due to the orientational change of the Mn3 spin.

We also examine the densities of the spin-orbit coupled states, indexed as \textbf{J}=\textbf{L}+\textbf{S} [Fig.\ \ref{Fig.3}d-e].  The splitting between Mn3~3\emph{d}$_{\frac{3}{2}}$ and 3\emph{d}$_{\frac{5}{2}}$ is enlarged when $\alpha$~$\approx$~30$^\circ$.  As the spin direction changes, more electrons go into 3\emph{d}$_{\frac{5}{2}}$, leading to more mixing with O$_{\text{--}}$~2\emph{p}.  These analyses provide an orbitally resolved understanding of how the charge density is redistributed through the Mn3-O bonds to drive the overall ferroelectric polarization along the [111] direction.

In summary, our DFT+\textit{U}+\textit{J}+SOC calculations demonstrate that CaMn$_{7}$O$_{12}$ adopts a noncollinear magnetic ground state, with Mn3 spins arranged in the noncentrosymmetric (30$^\circ$, 90$^\circ$) configuration. The resulting Berry phase polarization is nearly pure electronic with negligible Mn displacements.  According to the generalized spin-current model~\cite{Xiang11p157202}, the polarization is proportional to the sine of the Mn3 spin angle; it is coupled to the spin helicity, vanishing and reversing its direction at the centrosymmetric (60$^\circ$, 60$^\circ$) configuration.  The charge density redistribution along the Mn3-O bonds, as evidenced by our orbital-projected density of states, is understood in terms of the directionally inequivalent exchange interactions within the Heisenberg-exchange DM-interaction model.  DM interaction stabilizes (30$^\circ$, 90$^\circ$) over (60$^\circ$, 60$^\circ$) configuration, and the resulting inequivalence in symmetric exchange leads to a weak nonionic striction and a spontaneous electronic polarization.  Our findings suggest the existence of magnetically induced ferroelectricity in nearly inversion-symmetric ion lattice, opening the avenue for ultrafast magnetoelectric effect in a single ferroelectric-magnetic domain.

We thank A. Brooks Harris, Eugene J. Mele, and Charles Kane for fruitful discussions.  J. S. L. wishes to thank the Vagelos Integrated Program in Energy Research (VIPER) at the University of Pennsylvania.  D. S.-G. was supported by the Office of Naval Research under Grant No. N00014-12-1-1033.  A. M. R. was supported by the U.S. Department of Energy, under grant DE-FG02-07ER46431.  The authors acknowledge computational support from the High-Performance Computing Modernization Office (HPCMO) of the U.S. Department of Defense, as well as the National Energy Research Scientific Computing (NERSC) center.

\bibliography{R167}

\end{document}